# Design and fabrication guide for a new metamaterial as an absorber of visible light with exceptionally high absorption efficiency

Vivek Khichar[1], Nader Hozhabri[2*] and Ali R. Koymen[1]

[1]Department of Physics, University of Texas at Arlington, Arlington, Texas 76019

[2]Nanotechnology Research Center, Shimadzu Institute, University of Texas at Arlington, Arlington, Texas 76019

[*]Author to whom correspondence should be addressed: nh@uta.edu

**ABSTRACT**

We present a guide for the design and fabrication of a CMOS-compatible metamaterial microstructure as an absorber of visible light with exceptionally high absorption efficiency (~ 98%), for wavelengths 400nm-700nm. The structural parameters of the microstructure have been optimized by using Finite Element method (FEM) based COMSOL multyphysics software simulations. An optimized 2D unit cell of the structure consists of $4um \times 160\ nm$ TiN base on glass substrate covered with 70 nm thick silicon dioxide ($SiO_2$). A periodic structure of titanium nitride (TiN) straps (each of 90 nm thick and 2-micron wide) is deposited over $SiO_2$. The straps are capped with 40 nm thick layer of high-temperature dielectric hafnium dioxide ($HfO_2$) with periodicity of 4-micron. This unit is symmetric along the other dimension and repeated periodically along the horizontal direction. Additionally, though this microstructure was optimized for visible light spectrum, it also shows almost similar absorption of

~96% intigrated over wavelenght spectrum from 400 nm to 1200 nm. This investigation shows good agreement between simulation and experimental results.

There is significant recent interest in the development of broadband absorbers of sunlight.[1-16] For example, Liu et al. studied titanium nitride (TiN) metasurface and found that 250 nm-thick TiN exhibited broadband absorption over the ultraviolet-visible-near infrared range. Also, those researchers further observed that an array of TiN nano-resonators and $TiO_2$ coating structure cooperatively provide multiple resonant modes, which introduces strong coupling with the solar radiation and eventually produces an ultra-broadband absorption.[17] Additionally, the structure showed $\geq 91\%$ absorption for the visible spectrum, and performance was independent of the polarization and incident angles.[11]

Choudhry et al. conducted analytical and numerical investigations for a periodic one dimensional (1D) grating of a multilayered alternating metal-dielectric tapered structure and reported over 99% absorption using molybdenum - germanium layers at normal incidence in the visible and near-infrared regimes of the solar spectrum.[16]

Akafzade et al. introduced a metamaterial nanostructure consisting of $TiN/SiO_2/TiN-HfO_2$-disk for use as a broadband solar absorber approaching almost 100% absorption in the visible part of the solar spectrum.[1] Based on simulations, Akafzade and Sharma reported absorption efficiencies of almost 100% near 600 nm wavelength and ~ 98% integrated over the broadband spectrum from 250 nm to 1100 nm. The metamaterial nanostructure proposed in their article exhibits high absorption efficiency and high-temperature functionality due to the incorporation of a high-temperature dielectric hafnium dioxide $(HfO_2)$,[1, 19-29]. Hafnium dioxide is a suitable choice of dielectric for high temperature applications. For example, in comparison to the widely used dielectric in the semiconductor industries, silicon nitride ($Si_3N_4$) and hafnium dioxide ($HfO_2$) offer several useful properties.[30] Additionally, in comparison to $Si_3N_4$, $HfO_2$ is characterized by a dielectric constant that is 4-6 times higher than $Si_3N_4$ with optical transparency over a much wider range of wavelengths (250-2000 nm). Also its density in the solid state that is three times higher than $Si_3N_4$, with enhanced chemical stability, and a much higher melting point of 2758 $^0C$ (compared to 1900 $^0C$ melting point of $Si_3N_4$)[19, 21, 27, 29, 31,32-34].

Tiwari et al. have investigated the usefulness of $HfO_2$ dielectric properties for highly sensitive waveguide-coupled surface plasmon resonance sensors[19, 33].They have additionally used surface plasmon resonance measurements and computer simulations to study the efficacy of $HfO_2$ dielectric in Ag/$HfO_2$/Au waveguide-coupled multilayer structures.

In this study, we provide design and fabrication of a simpler metamaterial structure as a broadband solar absorber with extremely high absorption efficiency (~98%) for normal incidence and a broad range of incident wavelengths from 400 nm to 700 nm. In that regard, a set of devices consisting of 1-D periodic grating array of TiN and $HfO_2$ with pitch of 4 micron on $SiO_2$/ TiN film stack was fabricated and studied using IR spectroscopy.

For the proposed metamaterial strucure, the design parameters were optimized using Finite Element method (FEM) based COMSOL multyphysics software, simulations to provide the maximum possible absorber efficiency.[35] As required by COMSOL, a "physics controlled mesh" was chosen in which the COMSOL determines the concentration of mesh elements based on the size and refractive index of the materials in each domain, as shown in Figure-1.

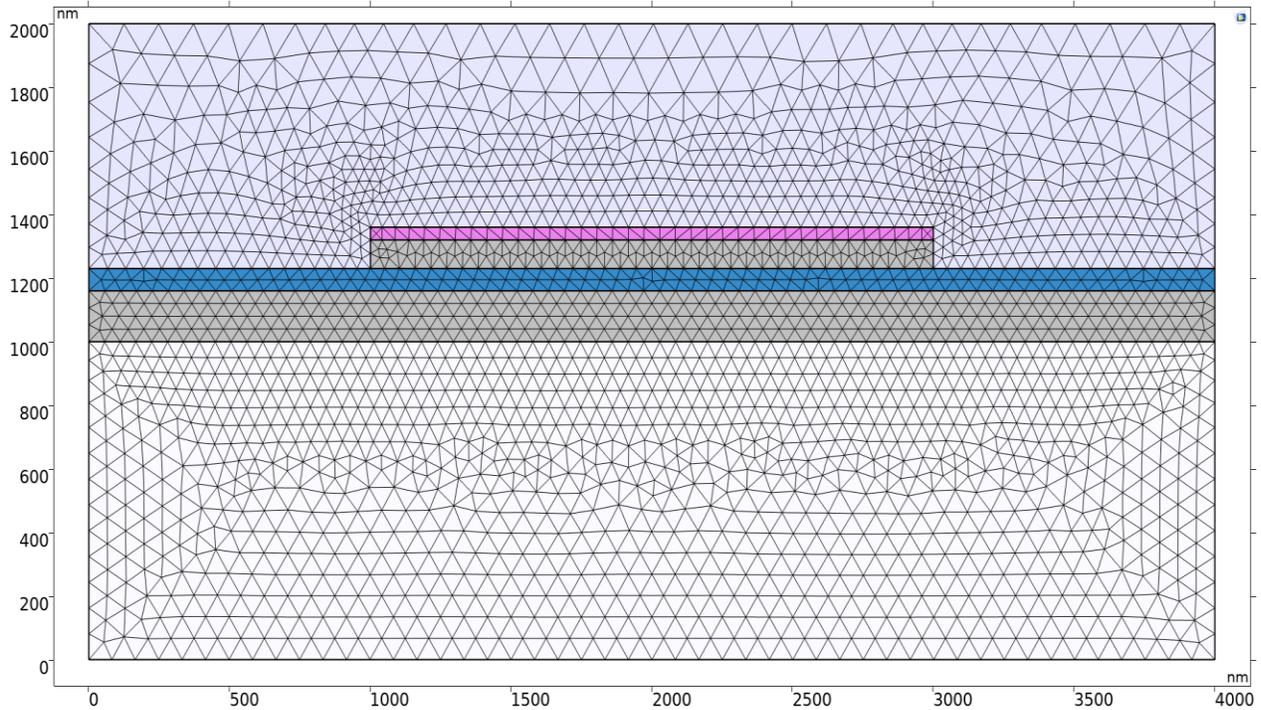

Figure-1. Physics controlled meshed structure of 4-micron pitch to simulate the absorption of incident light.

Simulation was carried out for visible light of 400 nm to 700 nm wavelength at normal incidece. All the thicknesses of the layers were optimized prior to manufacturing the device. The value of the optimized thicknesses for TiN/SiO$_2$/TiN/HfO$_2$ were found to be 160nm/70nm/90nm/40nm respectively..

The device was fabricated in Nanotechnology Research Center (NRC), a class-100 semiconductor fabrication facility. Figure-2 shows schematically the fabrication steps for the 4-micron pitch microstructure:

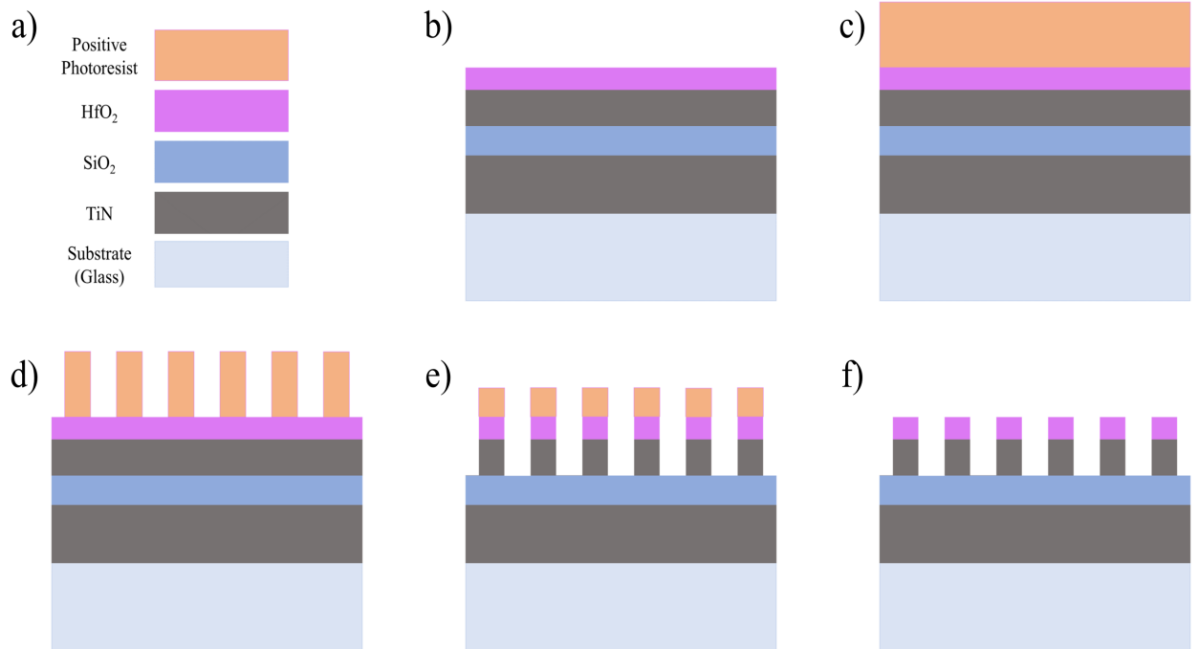

Figure-2. Schematics of the fabrication steps used for the metamaterial microstructure. a). color coding for used materials, b). deposited films stack on piranha cleaned substrate, c) coating of photoresist, d) post photolithography exposure, e) post etching and f) final structure after photoresist removal.

Device microfabrication began with the deposition of a 160 nm thick layer of TiN on piranha cleaned glass substrate using a multi gun AJA RF sputtering system. This layer was followed by deposition of a 70 nm thick $SiO_2$ and a 90 nm TiN film with the same sputtering equipment. The device was then transferred to a KJL Lab-18 RF sputtering deposition for deposition of a 40 nm $HfO_2$ thin film. KLA Tencor P6 Profilometer was used to measure the thickness of each deposited layer using simultaneously deposited layers on reference patterned silicon wafers.

Prior to pattering the stack through the photolithography process, the stack was dehydrated at 150°C for 30 minutes and cooled down followed by spreading monolayer of hexamethyldisilizane (HMDS) and baked at 150°C for 60 seconds. The film stack was then patterned by standard photolithography process using a custom designed photomask. For the patterning process, we used Shpley 1813 positive photoresist, and MF-319 developer.

Photolithography process steps are as follows: 1- Photoresist thickness:1.4μ, 2-soft bake (SB) at 125°C/ 60 seconds. 3-Exposure: 55mJ, 4-post exposure bake (PEB): 120°C/ 60 seconds, 5-development:75 seconds at room temperature (20°C).

Dry Etching: Stack films were etched in gas combination of CF4/Ar (10%/90%) at chamber pressure of 20 mTorr.

Post etching resist removal was carried out with Dow Microposit Remover 1165 at 70 °C.

Scanning Electron Microscopy (SEM) was used to inspect the grating profile of the microstructure and measure pitch, width of grating straps and trenches alongwith, for the validation of the structure fabrication process.

Figure-3 shows SEM image of patterned grating with measured pitch of about 4-micron with trench width of about 2-micron.

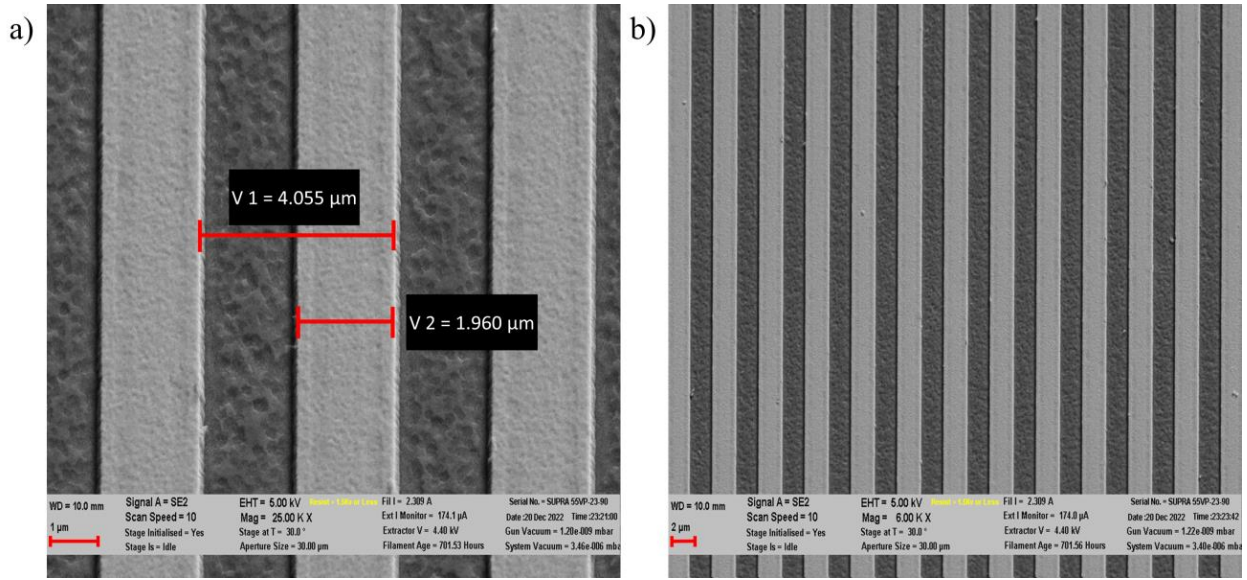

Figure-3. SEM images of microfabricated grating structures of 4-micron pitch. a) measurements of grating pitch and width of trench, b) density of grating walls.

Experimentally, absorption (A) data was collected using a V-570 UV-Vis Spectrometer by measuring reflectance (R) and transmittance (T) of microstructure for incident light of wavelength range from 400 nm to 1200 nm and wavelength step with 1 nm. Before R and T measurements, baseline corrections for the instrumet and background noise from any kind of source was measured to ensure the most accurate spectra measurements. Absorption of the device was calculated by using equation

$A = 1 - R - T$.

Figure-4 shows the experimental and simulated absorption of microfabricated structure of 4-micron grating pitch for mixed-polarized light at normal incidence as a function of wavelength. The integrated simulated absorption for the wavelength of 400nm to 1200nm is about 96% and about 94% for integrated experimental absorption.

On the other hand, for the visible range of wavelengths from 400nm to 700 nm, the integrated simulated absorption is about 98% and 94.3% for corresponding experimental absorption. The simulated and experimental results are in very good agrements. The less than 4% difference in the simulated and experimental values are attributed to the nature of complexity of the fabrication process such as films thickness variations, light diffraction, profiles of the structures that deviates from simulated vertical walls, and width of the lines from simulated 2μm as well as measurements errors.

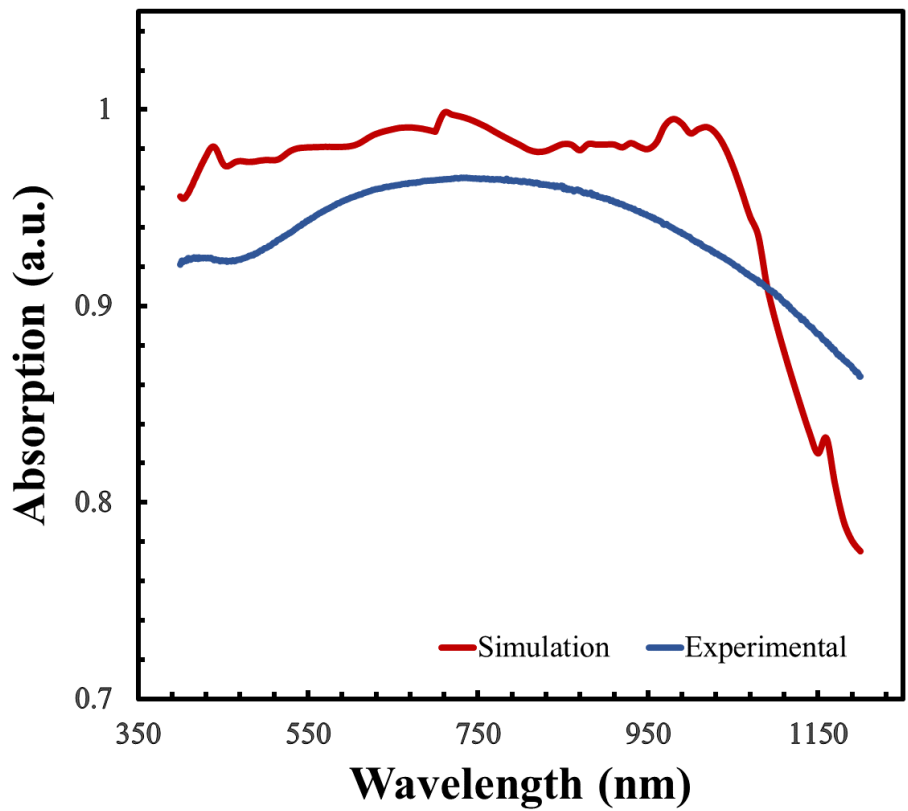

Figure-4. Absorption of 4-micron pitch grating metamaterial microstructure as a function of mixed polarized light wavelength at normal incident, where red curve shows simulated and blue curve shows experimental absorption data

## IV.    CONCLUSIONS

In conclusion, we have successfully investigated a new CMOS-compatible metamaterial microstructure as a potential broadband absorber of solar spectrum. The proposed metamaterial microstructure has distinct advantage that it can be fabricated readily, as well as easily by using standard semiconductor fabrication technology tools and high-performance materials; $SiO_2$, TiN, $HfO_2$. The structural parameters of the microstructure are optimized by using Finite Element method (FEM) based COMSOL multyphysics software. Subsequently, the absorption characteristics of the optimized and fabricated microstructure are evaluated as functions of the normal incident wavelength experimentally and using COMSOL. Additionally, the experimental studies of the performance of metamaterial microstructure shows good agreement with simulated absorption as broadband absorber and encouraging to fabricate nanostructure devices and investigate them experimentally. It is hoped that the design and fabrication guide presented in this work for a solar absorber will generate interest among the scientific and industrial communities involved in the research, development, and manufacturing of advanced and highly efficient solar panels.

# References

[1]Hussein Akafzade, S. C. Sharma. New Metamaterial as a Broadband Absorber of Sunlight with Extremely High Absorption Efficiency. AIP Advances.**10**:035209.(2020)

[2]G. Deng, K. Lv, H. Sun, Z. Yin, J. Yang. Stereo Perfect Metamaterial Absorber Based on Standing Gear-Shaped Resonant Structure With Wide-Incident-Angle Stability. Frontiers in Physics.**8**:609527.(2020)

[3]S. K. Patel, S. Charola, J. Parmar, M. Ladumor, N. Quang Minh, V. Dhasarathan. Broadband and efficient graphene solar absorber using periodical array of C-shaped metasurface. Optical and Quantum Electronics.**52**:250 (219 pp.).(2020)

[4]Q. Feng, C. Xifang, Y. Zao, Y. Weitang, Y. Hua, T. Yongjian, Y. Yong, L. Hailiang, Y. Yougen. Ultra-broadband and wide-angle perfect solar absorber based on TiN nanodisk and Ti thin film structure. Solar Energy Materials & Solar Cells.**211**:110535 (110510 pp.).(2020)

[5]S. Katsumata, T. Isegawa, T. Okamoto, W. Kubo. Effect of metamaterial perfect absorber on device performance of PCPDTBT:PC71BM solar cell. Physica Status Solidi A - Applications and Materials Science.**217**:1900910 (1900915 pp.).(2020)

[6]K. Xiangkun, J. Shunliu, K. Lingqi, W. Qi, H. Haobin, Z. Xiang, Z. Xing. Transparent metamaterial absorber with broadband radar cross-section (RCS) reduction for solar arrays. IET Microwaves, Antennas & Propagation.**14**:1580-1586.(2020)

7. Y. Peiqi, C. Xifang, Y. Zao, T. Yongjian, Y. Hua, Z. Zigang, D. Tao, C. Shubo, Z. Jianguo, Y. Yougen. A numerical research of wideband solar absorber based on refractory metal from visible to near infrared. Optical Materials. **97**:109400 (109406 pp.).(2019)

8. W. W. W. Leung, C. N. Savory, R. G. Palgrave, D. O. Scanlon. An experimental and theoretical study into NaSbS2 as an emerging solar absorber. J Mater Chem C. **7**:2059-2067.(2019)

9. R. Gudala, K. Songhee, K. Mee-Ree, K. K. Challa, S. Dong-Bum, V. K. Arepalli, K. Eui-Tae. Facile, cost-effective, nucleobase-mediated chemical deposition of solar absorber Cu2ZnSnS4 films. Applied Surface Science. **494**:756-762.(2019)

10. R. Harzallah, M. Larnicol, C. Leclercq, A. Herbein, F. Campana, editors. Development of high performances solar absorber coatings. SolarPACES 2018: International Conference on Concentrating Solar Power and Chemical Energy Systems, 2-5 Oct 2018; 2019; USA: AIP Publishing.

11. V. D. J. Herold, J. M. Dhavamani, D. K. Janapala. Step impedance resonator-based tunable perfect metamaterial absorber with polarization insensitivity for solar cell applications. Int J Rf Microw C E. **29**:e21650 (21657 pp.).(2019)

12. M. Bagmanci, M. Karaaslan, E. Unal, O. Akgol, M. Bakir, C. Sabah. Solar energy harvesting with ultra-broadband metamaterial absorber. International Journal of Modern Physics B. **33**:1950056 (1950016 pp.).(2019)

13. L. Herding, S. Caron, V. Nickich, F. Sutter, editors. Spectral characterisation of high temperature solar absorber coatings. 2018 7th International Energy and Sustainability Conference (IESC), 17-18 May 2018; 2018; Piscataway, NJ, USA: IEEE.

[14.]L. Zhengqi, L. Guiqiang, H. Zhenping, L. Xiaoshan, F. Guolan. Ultra-broadband perfect solar absorber by an ultra-thin refractory titanium nitride meta-surface. Solar Energy Materials & Solar Cells.**179**:346-352.(2018)

[15.]H. Akafzade. INVESTIGATION OF NANOSTRUCTURES FOR HIGHLY SENSITIVE SURFACE PLASMON RESONANCE , Ph. D. Dissertation (2020), Supervisor, Professor Suresh C Sharma: University of Texas at Arlington; 2020.

[16.]A. K. Chowdhary, D. Sikdar, editors. Ultra-broadband wide-angle metallo-dielectric metamaterial absorber for solar energy harvesting. 2019 Workshop on Recent Advances in Photonics (WRAP), 13-14 Dec 2019; 2019; Piscataway, NJ, USA: IEEE.

[17.]Z. Q. Liu, G. Q. Liu, Z. P. Huang, X. S. Liu, G. L. Fu. Ultra-broadband perfect solar absorber by an ultra-thin refractory titanium nitride meta-surface. Sol Energ Mat Sol C.**179**:346-352.(2018)

[18.]M. M. Bait-Suwailam, H. Alajmi, M. Masoud, editors. A Solar Energy Absorber Design Using Metamaterial Particles for Renewable Energy Solutions. 2nd IEEE Middle East and North Africa COMMunications Conference, MENACOMM 2019, November 19, 2019 - November 21, 2019; 2019; P.O. Box 5243, Manama, Bahrain: Institute of Electrical and Electronics Engineers Inc.

[19.]K. Tiwari, S. C. Sharma, N. Hozhabri. High performance surface plasmon sensors: Simulations and measurements. J Appl Phys.**118**.(2015)

[20.]T. Jõgiaas, M. Kull, H. Seemen, P. Ritslaid, K. Kukli, A. Tamm. Optical and mechanical properties of nanolaminates of zirconium and hafnium oxides grown by atomic layer deposition. Journal of Vacuum Science & Technology A.**38**:022406.(2020)

[21] V. Kolkovsky, K. Lukat, E. Kurth, C. Kunath. Reactively sputtered hafnium oxide on silicon dioxide: Structural and electrical properties. Solid-State Electron. **106**:63-67.(2015)

[22] T. M. Wang, C. H. Chang, J. G. Hwu. Enhancement of temperature sensitivity for metal-oxide-semiconductor (MOS) tunneling temperature sensors by utilizing hafnium oxide (HfO2) film added on silicon dioxide (SiO2). Ieee Sens J.**6**:1468-1472.(2006)

[23] R. A. Salinas Dominguez, A. Orduna-Diaz, S. Ceron, M. A. Dominguez. Analysis and Study of Characteristic FTIR Absorption Peaks in Hafnium Oxide Thin Films Deposited at Low-Temperature. Trans Electr Electro.**21**:68-73.(2020)

[24] B. Miao, R. Mahapatra, N. Wright, A. Horsfall. HfO2-based high-k dielectrics for use in MEMS applications. MEMS: Fundamental Technology and Applications: CRC Press; 2017. p. 21-39.

[25] M. N. Bhuyian, D. Misra. High- dielectrics and device reliability. Nano-CMOS and Post-CMOS Electronics: Devices and Modelling: Institution of Engineering and Technology; 2016. p. 1-33.

[26] T. Ando, U. Kwon, S. Krishnan, M. M. Frank, V. Narayanan. High-k oxides on Si: MOSFET gate dielectrics. Thin Films on Silicon: Electronic and Photonic Applications. 8: World Scientific Publishing Co. Pte. Ltd.; 2016. p. 323-367.

[27] M. Fadel, O. A. Azim, O. A. Omer, R. R. Basily. A study of some optical properties of hafnium dioxide (HfO2) thin films and their applications. Applied Physics a-Materials Science & Processing.**66**:335-343.(1998)

28. A. G. Bagmut, I. A. Bagmut, V. A. Zhuchkov, M. O. Shevchenko. Laser-deposited thin hafnium dioxide condensates: Electron-microscopic study. Technical Physics Letters.**38**:22-24.(2012)

29. J. P. Lehan, Y. Mao, B. G. Bovard, H. A. Macleod. Optical and Microstructural Properties of Hafnium Dioxide Thin-Films. Thin Solid Films.**203**:227-250.(1991)

30. K. Tiwari, S. C. Sharma, N. Hozhabri. Hafnium dioxide as a dielectric for highly-sensitive waveguide-coupled surface plasmon resonance sensors. Aip Advances.**6**.(2016)

31. H. F. Jiao, X. B. Cheng, J. T. Lu, G. H. Bao, Y. L. Liu, B. Ma, P. F. He, Z. S. Wang. Effects of substrate temperatures on the structure and properties of hafnium dioxide films. Appl Opt.**50**:C309-C315.(2011)

32. C. Adelmann, V. Sriramkumar, S. Van Elshocht, P. Lehnen, T. Conard, S. Gendt. Dielectric properties of dysprosium- and scandium-doped hafnium dioxide thin films. Appl Phys Lett.**91**.(2007)

33. K. Tiwari. Bimetallic waveguide-coupled sensors for tunable plasmonic devices: University of Texas at Arlington; 2015.

34. W. T. Liu, Z. T. Liu, F. Yan, T. T. Tan. Influence of RF power on the structure and optical properties of sputtered hafnium dioxide thin films. Physica B.**405**:1108-1112.(2010)

35. www.comsol.com. COMSOL Multiphysics. 2019. p. A general-purpose simulation software for modeling designs, devices, and processes in all fields of engineering, manufacturing, and scientific research.